# PPG2ABP: Translating Photoplethysmogram (PPG) Signals to Arterial Blood Pressure (ABP) Waveforms

Nabil Ibtehaz[1], Sakib Mahmud[2], Muhammad E. H. Chowdhury[*2], Amith Khandakar[2], Mohamed Arselene Ayari[3], Anas Tahir[4], M. Sohel Rahman[5]

[1]Department of Computer Science, Purdue University, West Lafayette, IN 47907, United States
[2]Department of Electrical Engineering, Qatar University, Doha 2713, Qatar
[3]College of Engineering, Qatar University, Doha 2713, Qatar
[4]Department of Electrical and Computer Engineering, University of British Columbia, Canada
[5]Department of CSE, BUET, ECE Building, West Palashi, Dhaka-1205, Bangladesh

*Corresponding author: Muhammad E. H. Chowdhury (mchowdhury@qu.edu.qa)

## Abstract
Cardiovascular diseases are one of the most severe causes of mortality, taking a heavy toll on lives annually worldwide. Continuous monitoring of blood pressure seems to be the most viable option, but this demands an invasive process, introducing several layers of complexities. This motivates us to develop a method to estimate the continuous arterial blood pressure (ABP) waveform through a non-invasive approach using Photoplethysmogram (PPG) signals. We explore the advantage of deep learning as it would free us from sticking to ideally shaped PPG signals only, by making handcrafted feature computation irrelevant, which is a shortcoming of the existing approaches. Thus, we present, PPG2ABP, a two-stage cascaded deep learning-based method, that manages to estimate the continuous ABP waveform from the input PPG signal, with a mean absolute error of 4.604 mmHg, preserving the shape, magnitude, and phase in unison. However, the more astounding success of PPG2ABP turns out to be that the computed values of Diastolic Blood Pressure (DBP), Mean Arterial Pressure (MAP), and Systolic Blood Pressure (SBP) from the estimated ABP waveform outperform the existing works under several metrics (mean absolute error of 3.449 ± 6.147 mmHg, 2.310 ± 4.437 mmHg and 5.727 ± 9.162 mmHg respectively), despite that PPG2ABP is not explicitly trained to do so. Notably, both for DBP and MAP we achieve Grade A in BHS (British Hypertension Society) Standard and satisfy the AAMI (Association for the Advancement of Medical Instrumentation) standard.

## Key Words
Blood Pressure, Cuff-less Blood Pressure, Mobile Health, Photoplethysmogram (PPG), Regression

## 1. Introduction

Even in today's world of technological advances, cardiovascular disease (CVD) is one of the most menacing causes of morbidity and mortality, crippling the ageing population [1]. More than 4 million people die of cardiovascular diseases every year only in Europe and when considering the whole world, the number of deaths exceeds 17 million [2]. Hypertension is one of the leading reasons for cardiovascular diseases. Alarmingly, only in the USA, there are around a million patients with hypertension, which cover almost one-third of the population, and shockingly only less than half of them try to control their blood pressure [3]. Hypertension has thus been termed a 'Silent Killer' due to its dormant nature that eventually leads to untimely death [4]. Therefore, continuous blood pressure monitoring is essential. However, owing to the lack of expert physicians, as opposed to the ever-increasing number of patients, the development of automated monitoring methods seems to be the only feasible means to confront the crisis.

Several methods have been introduced, capable of measuring blood pressure (BP) accurately. However, this accuracy comes at the cost of the invasiveness of such methods, which are often cumbersome to apply. For a catheter-based approach [5], not only the intervention of an expert is required, but also such procedures cause pain to already delicate patients. Clinics nowadays rely on cuff-based methods to measure blood pressure, but those as well are somewhat inconvenient and intrusive and more importantly, not suitable for continuous blood pressure monitoring [6]. Therefore, for a significant amount of time, it has been the interest of the research community to develop methods and apparatus to determine blood pressure from biomedical signals in a continuous, cuff-less, non-invasive manner [7]-[9]. Photoplethysmography (PPG) has gained a lot of popularity recently, due to its widespread inclusion in smartwatches and fitness bands, owing to its simplicity and cheapness. The idea behind PPG is rather simple; it works based on the illumination of skin and detection of the light absorption of skin. A PPG sensor generally comprises a Light Emitting Diode (LED) light source and a photodetector [10], the LED emits light to the skin tissue and the photodetector keeps track of the reflected light, i.e., the degree of absorption. It has been empirically established that the amount of reflected light is proportional to the volume of blood flowing in the illuminated region [11]. Since the volume of blood is related to the speed of blood flow which relates to the pressure exerted on the arteries, the PPG signal has been prominently used for the measurement of blood pressure [12]. Besides, PPG signals are also used for calculating the absorption of Oxygen as well as the level of Hemoglobin in blood [13] and for diagnosis of events like Hyperemia [14].

In recent years, a plethora of academic studies have been reported, to assess the state of blood pressure, using biomedical signals, mainly PPG often in conjunction with several other ones. The primary rationale behind measuring blood pressure from PPG is the association between the speed of blood flow and blood pressure. Overall, when blood vessels are contracted, blood flows rapidly, enforcing more pressure [15]. The opposite scenario is observed when the vessels are relaxed, steady blood flow diminishes the pressure. Therefore, studies have been conducted to investigate the rate of blood flow, which is popularly termed Pulse Wave Velocity (PWV) [16]. Based on PWV, blood pulses require a time delay to reach the periphery of the body from the heart, which is denoted as Pulse Transit Time (PTT) [17]. Two other parameters, namely, Pulse Arrival Time (PAT) and Pre-Ejection Period (PEP) are also relevant in such analyses [12]. A lot of research works have been conducted to develop mathematical models of these various parameters to infer blood pressure values [18]-[22]. While most prior works revolve around fitting the delay terms, lately, several classical machine learning-based approaches have been introduced [6], [15], [23]-[36]. These methods usually take the PPG signal, along with Electrocardiogram (ECG) signal in most cases and predict the values of Diastolic Blood Pressure (DBP), Systolic Blood Pressure (SBP), and Mean Arterial Pressure (MAP). Some other studies used Deep Machine Learning based approaches for BP prediction from PPG and/or ECG signals [37]-[46]. However, they are suffering from some limitations. Firstly, most of these methods require ECG signals alongside PPG, which may be difficult to include in wearable cuff-less systems. Secondly, some of these rely on handcrafted features to predict the BP; but to robustly compute such features, the algorithms often demand the signals to be of near ideal condition all the time, which is impractical. To solve these issues, in recent years, there have been a few studies which tried to estimate or reconstruct Arterial Blood Pressure (ABP) waveforms from PPG and/or ECG signals using various deep learning techniques such as 1D segmentation [47]-[48], variational autoencoders (VAEs) [49], and CycleGAN [50].

This work presents PPG2ABP, a pioneering framework in this domain that uses purpose-built, cascaded deep learning-based 1D segmentation models for continuously estimating the waveshape of the ABP signals from only PPG signals. The first segmentation network in the PPG2ABP pipeline tries to approximate the ABP waveform while the second network takes in the approximated ABP waveforms and refines them. Furthermore, being a deep learning-based pipeline, PPG2ABP is free from the need for handcrafted features, therefore the requirement of signals maintaining a standard shape is not essential. Moreover, the different values of interest in the literature, i.e., SBP, DBP, and MAP can be calculated from the estimated ABP waveform, and even in this objective, our method outperforms most of the existing works, despite not being explicitly trained to do so. The primary contributions of this study are as follows:

- To overcome the challenges in ABP estimation, we propose PPG2ABP which is a cascaded approach to divide this challenging task into two stages and reach a robust outcome in the end.
- The Approximation network approximates the ABP waveforms, and the Refinement network refines the outputs of the Approximation network.
- Our proposed PPG2ABP only requires PPG waveforms for ABP estimation, thus mitigating the need for ECG probes in parallel to PPG collection devices. This makes the solution simple, cost-effective, and user-friendly.
- PPG2ABP performs better than most studies in the literature while working on a large dataset.

The rest of the paper is organized as follows: Section 2 described the methodology followed by this study in detail including a brief discussion about the dataset, data preprocessing stages, and the proposed PPG2ABP pipeline. Section 3 contains the experimental setup and comparative evaluations against the prior works using various evaluation metrics. The outcomes of the experiments and ablation studies are provided in Section 4 along with interactive visualizations. Finally, the article is concluded with Section 0.

## 2. Materials and Methods

In this section, we discuss in detail the dataset used in this study and the proposed methodology. In the methodology section, we elaborate on the data preprocessing steps and the proposed PPG2ABP pipeline.

### *2.1 Dataset*

To develop and evaluate our proposed algorithm, the Multiparameter Intelligent Monitoring in Intensive Care (MIMIC-III) dataset from PhysioNet [51]-[52] has been used. This database contains simultaneous PPG and ABP signals. The sampling rate for all signals is 125 Hz, recorded with 8-bit precision. In this work, we utilized the data compiled from MIMIC-III by Kachuee et al. [53]-[54]. This compiled dataset is present in the University of California Irvine (UCI) Machine Learning Repository [55]. The data present in this repository is in a convenient form to analyze, and the raw signals are already pre-processed followed by their algorithm. For the sake of convenience, Kachuee et al. [54] ignored the signal episodes with too tiny or too large blood pressure values. They only considered signals with (60 mmHg ≤ DBP ≤ 130 mmHg) and (80 mmHg ≤ SBP ≤ 180 mmHg). In this study, we wanted to test our algorithm on a broader range of signals since a real-world application scenario might exhibit extremely small and high BP values. Therefore, we considered signals with DBP as low as 50 mmHg and SBP as high as 200 mmHg. The statistics of the dataset are presented in Table 1. It can be observed that SBP has a comparatively greater value of standard deviation; this extensive range is likely to cause severe difficulties when predicting SBP as hypothesized by Kachuee et al [54].

**Table 1** Statistics of the Dataset. Here we present the minimum, maximum and average values of DBP, MAP , and SBP respectively. In addition, we also list their standard deviation. It can be observed that SBP values have the highest variance, which makes their prediction the most difficult one, comparatively.

| | **Min** | **Max** | **Mean** | **Std** |
|---|---|---|---|---|
| **DBP** | 50 | 165.17 | 66.14 | 11.45 |
| **MAP** | 59.96 | 176.88 | 90.78 | 14.15 |
| **SBP** | 71.56 | 199.99 | 134.19 | 22.93 |

For our analysis, we considered signal episodes of $T_e$ = 8.192 seconds long, i.e., we estimated $T_e$ seconds long arterial blood pressure (ABP) waveform from PPG signals of $T_e$ seconds long. Picking this $T_e$ of 8.192 seconds (which translates to 1024 samples), allowed us to train a sufficiently deep neural network without being crippled by extensive computational complexity. However, in the downloaded database, we have signals of a total duration of 741.53 hours. To mitigate the computational requirements, we used a portion of the available data. The following scheme was followed. First of all, the $T_e$ seconds-long signals were arranged in bins based on their SBP and DBP values. Next, from all the bins, 25% of the data was selected randomly. However, if for some bins selecting one-fourth exceeds 2500 episodes, then only 2500 episodes are included randomly. In this way, a total of 127260 random episodes were obtained, counting up to 353.5 Hours. From these 100000 samples at random were chosen to be the training data (roughly 78.58%) and the remaining 27260 samples were kept as the independent testing data. It should be noted that proper care and attention were given to make sure signal data from all the patients were included and also that the training and test data remained disjoint. Furthermore, we did not omit PPG signals of sub-ideal quality; rather the random selection process led to the inclusion of a high number of merely acceptable and unfit signals [56].

### *2.2 Proposed Methodology*

The proposed algorithm PPG2ABP extracts PPG segments of $T_e$ seconds long performs some minimal preprocessing (e.g., filtering) operation to attenuate the irregularities and noises. Next, the filtered signal is processed using an Approximation Network that approximates the ABP waveforms based on the input PPG signals. The preliminary rough estimate of ABP is further refined through a Refinement Network. Finally, in addition to the estimated ABP waveform, the values of SBP, DBP, and MAP are computed in a simple manner. The overall pipeline of PPG2ABP is depicted in Fig. 1.

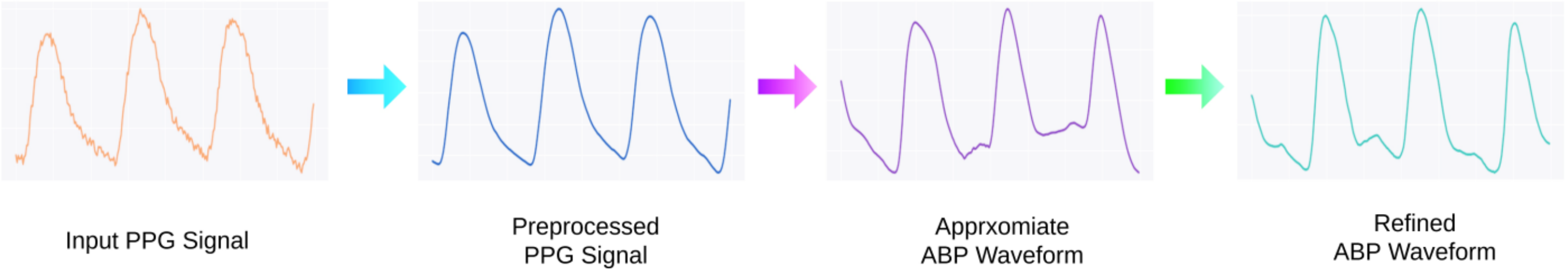

**Fig. 1.** Algorithmic Pipeline of PPG2ABP. PPG2ABP takes PPG segments of $T_e$ seconds long as the input and performs some preprocessing [53]. Next, the preprocessed signal is passed to the Approximation Network to approximate the ABP waveform. After that, the Refinement Network refines the approximated ABP waveforms. Finally, in addition to the ABP waveform, BP values such as SBP, MAP, and DBP are computed.

### *2.2.1 Preprocessing*

As mentioned earlier, for this study, we have used the signals already pre-processed by Kachuee et al. [53]. Therefore, our preprocessing steps are almost identical to theirs, except for some additional steps to prepare the dataset suitable for the deep learning pipeline. The preprocessing stage primarily involves wavelet denoising [54]. The wavelet transform is performed to 10 decomposition levels, with Daubechies 8 (db8) as the mother wavelet [57]. Then, the very low (0-0.25 Hz) and very high frequency (250-500 Hz) components are negated by setting the decomposition coefficients to zero, followed by soft Rigrsure thresholding [58], [59]. Finally, the signal is retrieved by reconstructing the decomposition. To facilitate the training of the deep learning models, the PPG signals were segmented, filtered, and global min-max normalized. After reconstruction, the ABP waveforms are denormalized using the global min-max values obtained earlier during normalizing.

### *2.2.2 Approximation Network*

The filtered signals are processed through an Approximation Network, that approximates the ABP signal. The Approximation Network is a one-dimensional deeply supervised U-Net model. U-Net [60] has been primarily constructed using only convolutional layers to perform the task of semantic segmentation. The network structure consists of symmetric pairs of encoder-decoder layers. The most innovative idea behind U-Net is the use of skip connections to preserve the spatial feature maps lost during pooling and upsampling.

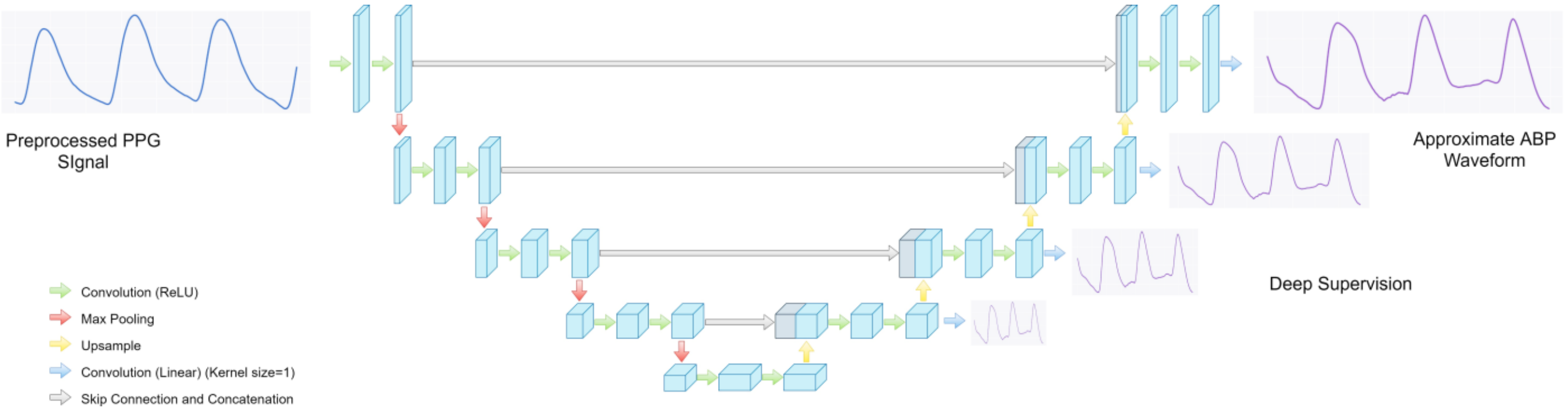

**Fig. 2.** Approximation Network. PPG2ABP employs the Approximation Network to derive an approximate ABP waveform from the preprocessed PPG signal. The network is a deeply supervised 1-dimensional U-Net, designed to solve regression problems.

Though the original U-Net is designed to perform semantic segmentation on images, for our purpose, we employ it to reconstruct 1-dimensional signals, which is primarily a regression task. Therefore, the two-dimensional convolution, pooling, and upsampling operations are replaced by their one-dimensional counterparts. To produce a regression output, the final convolutional layer uses a linear activation function. Moreover, we apply deep supervision in our U-Net network [61]. Deep supervision is a technique proven to reduce overall errors by guiding the learning process of the hidden layers. In our deeply supervised 1D U-Net, we compute an intermediate output which is a subsampled version of the actual output signal, before every upsampling operation in the decoder. The losses are computed with gradually declining weights as we move deeper into the model. These auxiliary losses drive the training of the hidden layers and make the final output much superior. The diagram of the Approximation Network has been presented in Fig. 2.

### 2.2.3 Refinement Network

The output of the Approximation Network sometimes deviates much from the target. Therefore, we use an additional network, namely 'Refinement Network', to refine the output of the Approximation Network. The Refinement Network is a 1-dimensional MultiResUNet model [62], an improved version of the U-Net model. The primary distinction between the two is the inclusion of the Multi-Residual or MultiRes blocks and the Residual or Res paths in the MultiResUNet model. The MultiRes blocks involve a compact form of multiresolution analysis using factorized convolutions. On the other hand, Res paths impose additional convolutional operations along the shortcut connections to reduce the disparity between the feature maps of the corresponding levels of encoders and decoders.

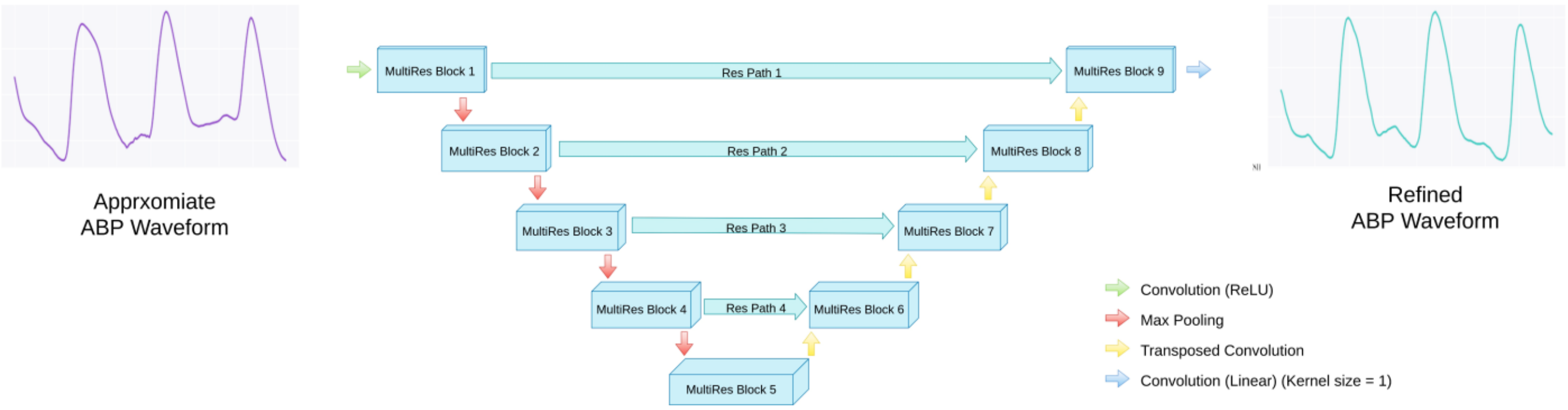

**Fig. 3.** Refinement Network. PPG2ABP pipeline utilizes a 1-dimensional MultiResUNet Network to refine the approximate waveform constructed by the Approximation Network.

Similar to the Approximation Network, this network comprises one-dimensional versions of convolution, pooling, and upsampling operations, and the final convolutional layer uses linear activation. However, this model is not deeply supervised. The diagram of the Refinement Network is presented in Fig. 3.

### 2.2.4 BP Parameters Calculation

In addition, to construct the continuous ABP waveforms, the typical BP values of interest, namely, SBP, DBP, and MAP can be computed by taking the max, min and mean values of the estimated signals, respectively as shown in Eqs. (1) to (3).

$$SBP = max(ABP) \tag{1}$$

$$DBP = min(ABP) \tag{2}$$

$$MAP = mean(ABP) \tag{3}$$

Here, ABP is the estimated Arterial Blood Pressure waveforms from the Refinement Network.

## 3. Experiments

In this section, we reflect on various aspects of our experimental setup and insights into the ablation studies performed to determine the optimum parameters for the deep learning architectures to estimate ABP waveforms from PPG signals. For this study, we have used the Python programming language to implement the algorithms and conduct experiments. The neural network models have been developed using TensorFlow 2.0 with a KERAS frontend. Moreover, we have made the codes publicly available, which can be found in [63].

### 3.1 Selection of Models

In addition to U-Net and MultiResUNet, we also conducted some preliminary experiments using other deep learning models. However, U-Net and MultiResUNet yielded better results. We also experimented with the combinations of U-Net and MultiResUNet for the approximation and refinement networks. It was observed that U-Net as the refinement network, failed to reach the performance level of MultiResUNet. On the contrary,

MultiResUNet as the approximation network performed better than U-Net, but the overall performance followed by a MultiResUNet refinement network remained quite identical. We hypothesize that through MultiResUNet is superior to U-Net and manages to obtain a much better waveform, the refinement network reaches a plateau eventually. Nonetheless, since U-Net is computationally lighter than MultiResUNet, we use U-Net and MultiResUNet as the approximation and the refinement network, respectively.

### *3.2 Selection of Loss Functions*

Mean Squared Error (MSE) and Mean Absolute Error (MAE) is the most prevalently used loss functions for regression. For predicted values $\widehat{\mathrm{Y}} = [\hat{\mathrm{y}}_1, \hat{\mathrm{y}}_2, \hat{\mathrm{y}}_3, \ldots, \hat{\mathrm{y}}_\mathrm{n}]$ and the ground truth $\mathrm{Y} = [\mathrm{y}_1, \mathrm{y}_2, \mathrm{y}_3, \ldots, \mathrm{y}_\mathrm{n}]$, they are defined in Eqs. (4) and (5) as follows:

$$MSE = \frac{\sum_{i=1}^{n} (y_i - \hat{y}_i)^2}{n} \quad (4)$$

$$MAE = \frac{\sum_{i=1}^{n} |y_i - \hat{y}_i|}{n} \quad (5)$$

In our experiments, we found that using MAE as the loss function of the approximation network (as opposed to MSE) significantly improves the performance. Upon inspecting samples and outputs, we developed the following rationale. Since at the approximation network stage we aim to obtain a rough estimate of the waveform, it suffices to put equal weights to the entire range of errors. But MSE squares the error terms, and the bigger errors are more penalized. At this stage, we have rather little information regarding the output waveform; therefore, putting more emphasis on eliminating the bigger error terms degrades the overall performance. Thus, MAE in the approximation network stage balances all the error terms, ensuring a rough yet satisfactory projection. On the contrary, in the refinement network, we already have an approximation of the waveform. Hence, it becomes beneficial to use MSE in that stage as the larger error terms will get diminished better. The empirical evidence also supports this.

### *3.3 Effect of Number of Convolutional Filters*

We have explored a pool of wider variants of the networks, comprising an increasing number of convolutional kernels or filters. From experiments, it was observed that the models with a higher number of filters performed better, which is obvious since the inclusion of additional filters would allow the model to learn and capture additional shapes and features. However, as the number of filters increases, the models become exponentially heavier and computationally more expensive. So, after a certain level, the improvement obtained from adding new filters is not worth the rising computational demand. Therefore, we have used a U-Net model with the number of filters as multiples of 64, i.e., [64, 128, 256, 512, 1024] and for the MultiResUNet, we have limited the value of weight multiplier alpha ($\alpha$) to 2.5, a parameter which controls the number of filters [62]. For both our approximation and refinement networks, we have used the standard configurations of the convolutional and pooling layers. In the convolutional layers, filter or kernel size was selected to be 3, while stride length was 1. Additionally, padding was used to zero pad the intermediate outputs to keep them consistent with the input shape. On the other hand, for max-pooling layers, we used both pool size and stride length of 2. Similarly, in the upsampling layers, a window size of 2 was selected.

### *3.4 Effect of Deep Supervision*

Additionally, we have experimented with the concept of deep supervision and employed auxiliary losses during training. For both U-Net and MultiResUNet models, we have imposed additional loss functions on the outputs of the convolutional layers just before the transposed convolution (i.e., upsampling through convolution) operations. Moreover, the weights of the losses have been selected as [1, 0.9, 0.8, 0.7, 0.6], i.e., the maximum weight has been put on the final output and is gradually diminished for the premature outputs from deeper decoder layers. For the U-Net model, a dramatic improvement was observed after applying deep supervision while for the MultiResUNet model, the improvement was minimal. Therefore, to establish a trade-off between computational effort and accuracy, deep supervision has been employed in the U-Net model (i.e., the Approximation Model) only.

### *3.5 Training Methodology*

As specified in Section 3.2, MAE and MSE are used as the loss functions of the Approximation and the Refinement networks, respectively. To minimize these losses, the Adam optimizer [64] is used, which adaptively computes different learning rates for individual parameters based on the estimates of the first and second moments of the gradients. In our experiments, we used Adam with the default parameters mentioned in the original paper [64]. Each model has been trained for 100 epochs with a patience of 20.

### *3.6 K-Fold Cross Validation*

We have performed a 10-fold cross-validation using the training data, i.e., 90% of the training data is used to train the model and the remaining 10% data is used for validation. This approach is repeated 10 times using different data splits, and thus 10 models are developed. The best-performing model is selected and is then evaluated against the independent test set.

### *3.7 Evaluation Metrics*

Since we are primarily working with a regression problem, we have used Mean Absolute Error (MAE) defined in Eq. (5) as our primary evaluation metric. Furthermore, we have evaluated the outcomes from the PPG2ABP pipeline using domain-specific metrics such as the British Hypertension Society (BHS) Standard, Association for the Advancement of Medical Instrumentation (AAMI) Standard, Pearson Correlation Coefficient (PCC), Bland Altman plots, etc. The details of these metrics have been presented in the following sections.

## 4. Results

After training the PPG2ABP model, we evaluate the pipeline on the test data. The following outcomes have been derived from the evaluation of the model on the independent test set.

### *4.1 Estimating ABP Waveform*

The primary and unique objective of this work is to transform PPG signals into the corresponding ABP waveforms. Despite some correlation between the two as established from the past studies [15], [23], they are structurally quite different from each other when we consider the two waveforms. Nevertheless, the proposed PPG2ABP model manages to estimate the waveform of blood pressure by taking only the PPG signal as input. The output of the approximate network gives an overall rough estimate, which is further refined by the refinement model. Fig. 4 illustrates such an example. It can be seen that although the output from the Approximate Network follows the rough silhouette, it fails particularly in the rapid slope down from the peaks. However, after refining this waveform using the Refinement Network, such erroneous constructions are revised and improved, closely mimicking the ground truth. For this particular example, the use of the Refinement Network reduces the mean reconstruction error from 9.52 mmHg to 2.37 mmHg. The constructed waveform closely follows the ground truth waveform. Therefore, from experimental results it is evident that PPG2ABP can translate PPG signals to corresponding blood pressure waveforms, preserving the shape, magnitude, and phase in unison. Quantitatively, the mean absolute error (MAE) of this blood pressure waveform construction is $4.604 \pm 5.043$ mmHg over the entire test dataset. In addition, the mean absolute error of DBP, MAP, and SBP prediction is $3.449 \pm 6.147$ mmHg, $2.310 \pm 4.437$ mmHg, and $5.727 \pm 9.162$ mmHg, respectively. Furthermore, previous studies have pointed out that there exists a phase lag between the PPG and ABP signals of the MIMIC-III database [65], and some further processing is required to align them. However, in our generated output, we can observe that our method has been able to remarkably overcome this issue of phase lag. Indeed, this may turn out to be highly beneficial in dealing with the phase lag between the two signals in real-world applications due to the difference in Pulse Arrival Time (PAT) for recording them from two separate body locations. So, unlike existing works, we do not exclude irregular signals (details in the **supplementary materials**).

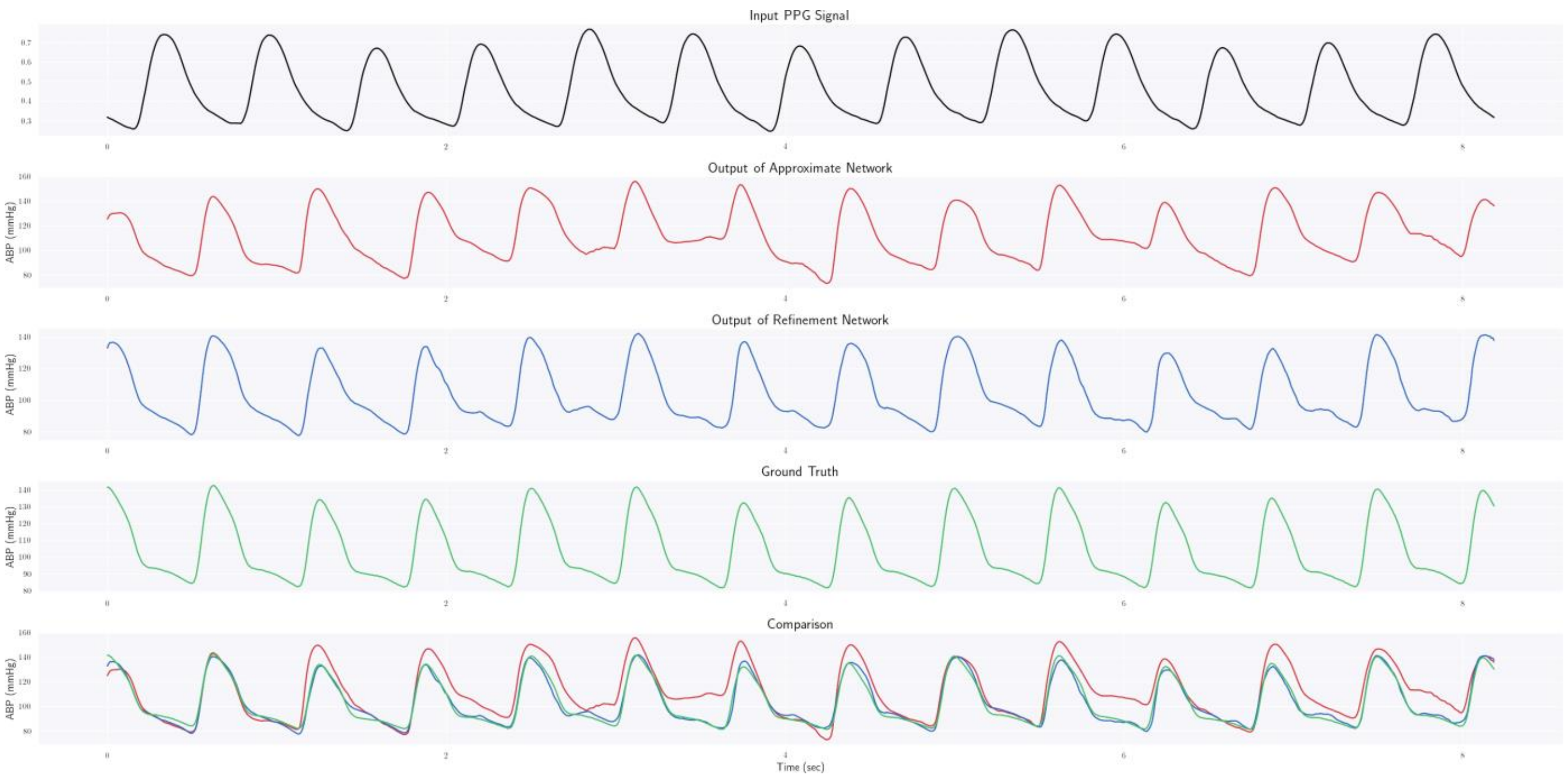

**Fig. 4.** Demonstration of output from the PPG2ABP pipeline

A sample PPG signal is shown in Fig. 4, from the test data is given as input. It can be observed that the output from the Approximation Network, despite roughly following the overall pattern of the ground truth, falls short in certain aspects, vividly apparent around the peaks. Furthermore, the reconstructed waveform fails to rapidly slope down from the peak regions. However, the estimation from the Refined Network seems to be more satisfactory. It can be observed that in addition to following the overall pattern of the ground truth waveform, the final estimated waveform also successfully mimics the peak regions and subsequent downward inclination. Therefore, the inclusion of the Refinement Network on top of the Approximate Network significantly improves the results, as evident from the drop of mean reconstruction error from 9.52 mmHg to 2.37 mmHg in this particular example.

### *4.2 BHS Standard*

The British Hypertension Society (BHS) introduced a structured protocol to assess blood pressure measuring devices and methods [66]. Hence, this standard has been frequently used in the literature as a metric [6], [23], [53]-[54]. The accuracy criteria of the BHS standard appraise methods based on absolute error. The grades are provided by counting what percentage of the predictions on the test samples fall under 5 mmHg, 10 mmHg, and 15 mmHg absolute error for grades A, B, and C, respectively. The three grades are presented in Table 2. For an algorithm to obtain a certain grade, it must satisfy all three thresholds simultaneously. In addition, there is a grade D for algorithms failing to meet the requirements of grade C [66].

**Table 2** Evaluation of BHS Standard. Here we present the criteria used in grading the rank of predictions using the BHS Standard. We also demonstrate how our results compare with the BHS Standard.

| | | **Cumulative Error Percentage** | | |
|---|---|---|---|---|
| | | $\leq$ 5 mmHg | $\leq$ 10 mmHg | $\leq$ 15 mmHg |
| **Our Results** | DBP | 82.836 % | 92.157 % | 95.734 % |
| | MAP | 87.381 % | 95.169 % | 97.733 % |
| | SBP | 70.814 % | 85.301 % | 90.921 % |
| **BHS** | Grade A | 60 % | 85 % | 95 % |
| | Grade B | 50 % | 75 % | 90 % |
| | Grade C | 40 % | 65 % | 85 % |

The absolute error of computing DBP, MAP, and SBP is presented in Fig. 5. We have presented how the absolute error of predicting DBP, SBP, and MAP of the samples are distributed, with particular focus on the error boundaries of 5 mmHg, 10 mmHg, and 15 mmHg. We have achieved grade A for both DBP and MAP, but grade B for SBP. To the best of our knowledge, the proposed algorithm is a pioneer work and one of the first that obtained a grade B in SBP prediction on a significant portion of data from the MIMIC-III dataset (more details in Section 4.5).

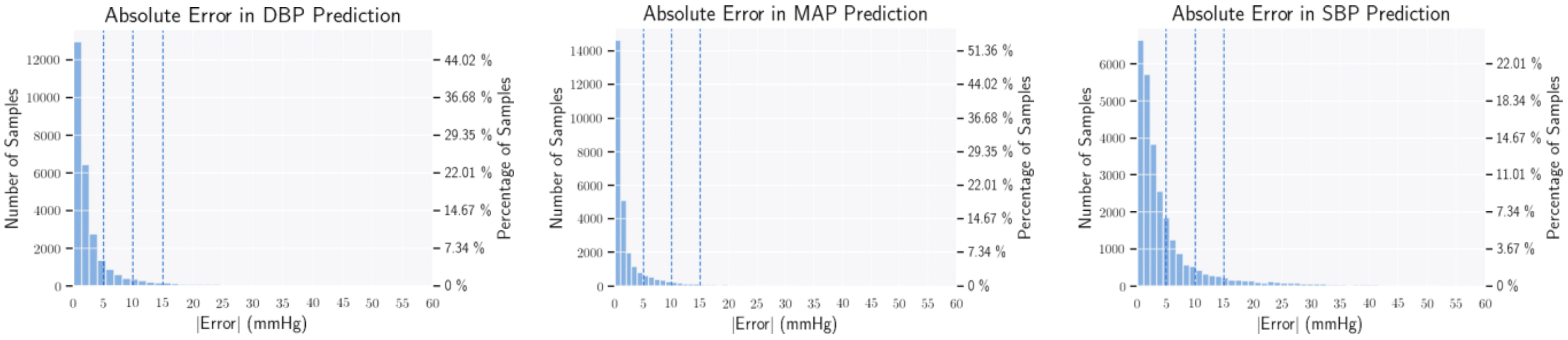

(a) (b) (c)

**Fig. 5.** Mean Absolute Error Histogram. Here we present how the absolute error of predicting DBP, SBP, and MAP of the samples are distributed. In addition, we also observe errors in how many samples lie below the 5 mmHg, 10 mmHg, and 15 mmHg thresholds, used in the evaluation of the BHS Standard.

### *4.3 AAMI Standard*

Similar to the BHS Standard, the Association for the Advancement of Medical Instrumentation (AAMI) Standard is another metric to evaluate blood pressure measuring devices and methods. The criterion set by the AAMI standard [67] requires the blood pressure measuring methods to have a mean error and standard deviation of less than 5 mmHg and 8 mmHg, respectively and the minimum number of subjects under study is 85. Table 3 shows our results under the AAMI criterion.

**Table 3** Evaluation of AAMI Standard. Here we present the criterion used in grading the rank of predictions using the AAMI Standard. We also demonstrate how our results compare with the AAMI Standard.

| | | **ME (mmHg)** | **STD (mmHg)** | **Number of Subjects** |
|---|---|---|---|---|
| **Our Results** | DBP | 1.619 | 6.859 | 942 |
| | MAP | 0.631 | 4.962 | |
| | SBP | -1.582 | 10.688 | |
| **AAMI Standard** | | ≤ 5 | ≤ 8 | ≥ 85 |

It can be observed that for both DBP and MAP the requirements of the AAMI standard are satisfied quite convincingly. However, for SBP, although the condition of mean error is fulfilled, the value of standard deviation is a bit higher. It may be noted here that other contemporary methods fail to satisfy the AAMI criterion for SBP on the MIMIC-III dataset as well. The histograms of error for prediction of DBP, MAP, and SBP are presented in Fig. 6. It is apparent that both for DBP and MAP these errors have a mean of almost zero with a small value of standard deviation. For SBP the standard deviations seem more distributed.

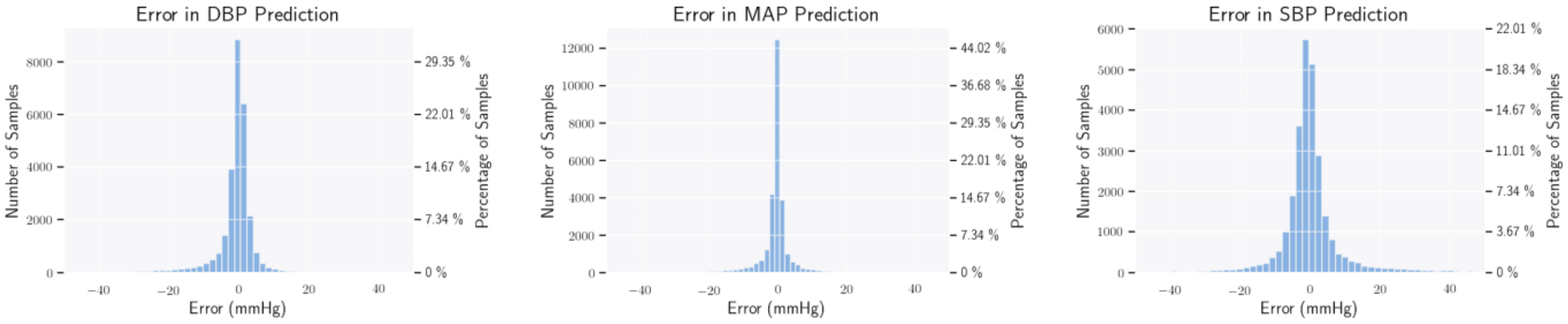

(a) (b) (c)

**Fig. 6.** Mean Error Histogram. Here, we show how the MAE of DBP, SBP, and MAP of the estimated samples are distributed. From the plots, all errors have a mean of zero and a small standard deviation, except SBP.

*4.4 Statistical Analysis*

Here, from Fig. 7, it is evident that the error of predicting DBP, MAP, and SBP of 95% of the samples lies between [-11.825:15.0637], [-9.095:10.357], and [-22.531:19.367], respectively. presents the Bland-Altman plots [68] for predicting DBP, MAP, and SBP respectively. The 95% limits of agreement span the segment from $\mu - 1.96\sigma$ to $\mu + 1.96\sigma$ (shown using dashed lines), where $\mu$ and $\sigma$ are the mean and standard deviation of the distribution, respectively. For DBP, MAP, and SBP, this limit translates to [−11.825:15.064], [−9.095:10.357], and [−22.531:19.367] mmHg, respectively. Though these numbers appear to be overwhelming, actually from the plots from Fig. 7 it can be seen that most of the error terms fall within the 5-mmHg range. Nevertheless, all three plots contain a great chunk of outliers, especially the SBP plot (Fig. 7c).

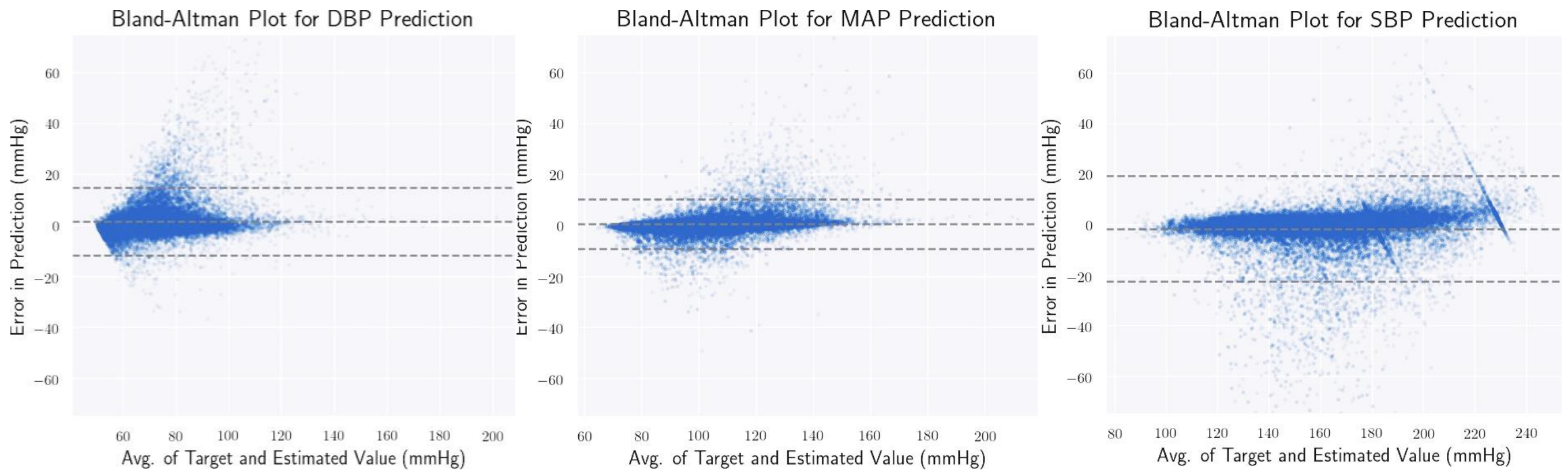

(a) (b) (c)

**Fig. 7.** Bland-Altman Plot. Here through the Bland-Altman plots, it is evident that the error of predicting DBP, MAP, and SBP of 95% of the samples lies between [-11.825:15.0637], [-9.095:10.357], and [-22.531:19.367], respectively.

In addition, Fig. 8 depicts the regression plots for predicting DBP, MAP, and SBP, respectively. From the plots, the correlation between the predictions and the ground truth is evident. Moreover, the values of the Pearson Correlation Coefficient (PCC) for DBP, MAP, and SBP predictions are 0.894, 0.966, and 0.936 respectively, indicating a strong positive correlation. Furthermore, such high values of PCC on a sample size of 27260 correspond to ($p < .000001$). Stating our null hypothesis to be the lack of any correlation between ground truth and predicted BP values, such a low p-value rejects the null hypothesis with a great margin, indicating the statistical significance of our results.

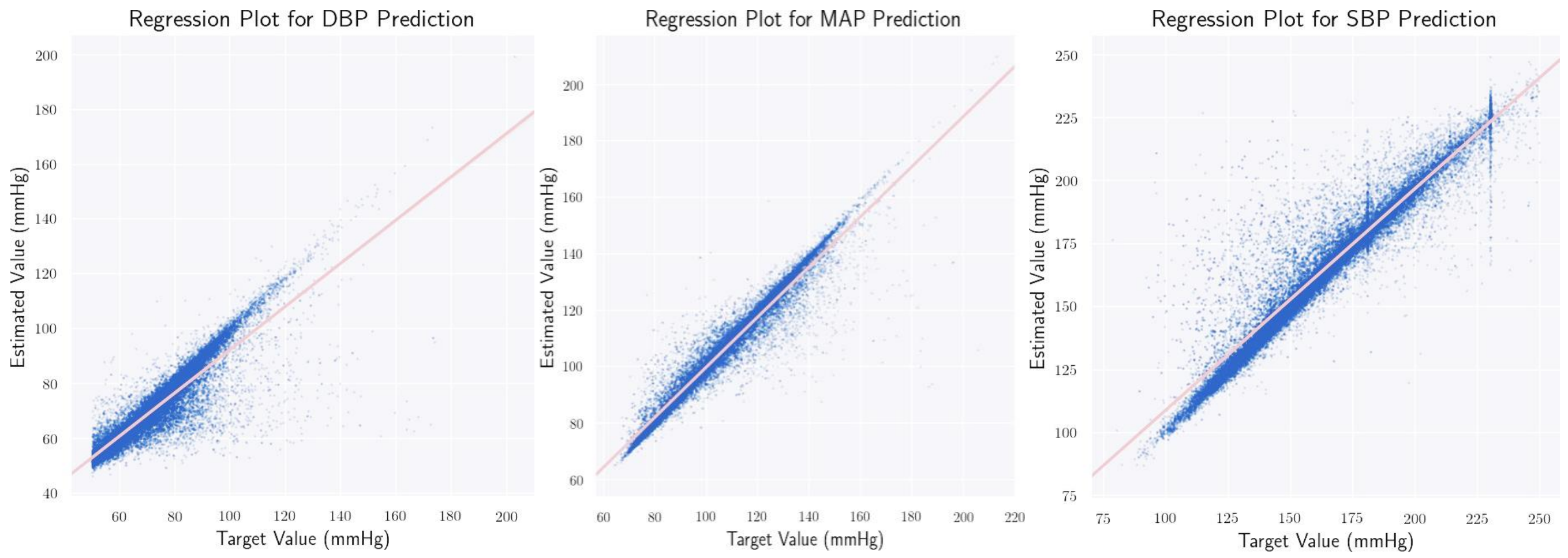

(a) (b) (c)

**Fig. 8.** Regression Plots for DBP, MAP, and SBP predictions, respectively. In all three cases, we obtain a p-value in the range of $p < .000001$, which rejects the null hypothesis and strengthens the statistical significance of our method.

### *4.5 Comparison with the Existing Methods*

Despite there being a lot of research endeavours on this topic, there are only some recent works which can be directly compared with our work. However, those recent articles cited our article when it first appeared in the archive. Therefore, this paper is the pioneering work on PPG to ABP synthesis. All other works used the concept of our work while our work was in the review phase for around two years before we withdraw the paper to submit here for fast review. However, to be fair in our comparison to the existing published literature, we have compiled a list of works evaluated on the MIMIC-III dataset with a comparable and sufficiently large number of patients and have presented a comparative analysis in **Table 4**. Since we have made the pre-processed data and code available on GitHub, several groups have worked on this dataset and managed to publish their article before us. A quote from the Qin et al. [49] paper with proper references, "*Ibtehaz et al. [69] firstly applies UNet [60], a classic network in the field of medical imaging, in conjunction with the concept of deep supervision, for generating ABP waveform from PPG signal. After that, Athaya et al. [47] did almost the same work as Ibtehaz et al. [69] based on UNet. Sadrawi et al. [70] applies two networks—LNet and UNet for building ABP waveform predictive model, respectively*." Therefore, this is the pioneering work in generating ABP signal from PPG alone, which has drawn significant attention from the research community.

**Table 4** Comparison among different approaches. Here we list the methods that used the MIMIC-III dataset to evaluate their performance. Furthermore, for a fairer comparison, we have only included the methods that consider a significant portion of the dataset. We compare the methods using measures like Mean Absolute Error (MAE) for predicting DBP, MAP, and SBP, in addition to BHS and AAMI Standard.

| **Study** | **Appearing Year** | **Dataset** | **Input** | **Results** |
|---|---|---|---|---|
| Kachuee et al. [53] | 2015 | MIMIC-III | PPG, ECG | BHS Standard: DBP = Grade B, MAP = Grade C, SBP = Grade D<br>MAE: DBP = 6.34 mmHg, MAP = 7.52 mmHg, SBP = 12.38 mmHg |
| Kachuee et al. [54] | 2016 | MIMIC-III | PPG, ECG | BHS Standard: DBP = Grade B, MAP = Grade C, SBP = Grade D<br>AAMI Standard met for DBP, MAP<br>MAE: DBP = 5.35 mmHg, MAP = 5.92 mmHg, SBP = 11.17 mmHg |
| Mousavi et al. [23] | 2019 | MIMIC-III | PPG | BHS Standard: DBP = Grade A, MAP = Grade B, SBP = Grade D<br>AAMI Standard met for DBP, MAP |
| Slapnivcar et al. [15] | 2019 | MIMIC-III | PPG | MAE: DBP = 9.43 mmHg, SBP = 6.88 mmHg |
| Athaya et al. [47] | 2021 | MIMIC-III | PPG | BHS Standard: DBP = Grade A, MAP = Grade A, SBP = Grade A<br>MAE: DBP = 2.17 mmHg, MAP = 1.97 mmHg, SBP = 3.68 mmHg |
| Harfiya et al. [48] | 2021 | MIMIC-III | PPG | BHS Standard: DBP = Grade A, MAP = Grade A, SBP = Grade A<br>MAE: DBP = 2.41 mmHg, SBP = 4.05 mmHg |
| Qin et al. [49] | 2021 | MIMIC-III | PPG | BHS Standard: DBP = Grade A, MAP = Grade A, SBP = Grade B<br>MAE: DBP = 7.95 mmHg, MAP = 3.83 mmHg, SBP = 4.11 mmHg |
| Mehrabadi et al. [50] | 2022 | MIMIC-III | PPG | BHS Standard: DBP = Grade A, MAP = Grade A, SBP = Grade A |
| | | | | MAE: DBP = 1.93 mmHg, SBP = 2.29 mmHg |
| **PPG2ABP (Proposed)** | 2020 | MIMIC-III | PPG | BHS Standard: DBP = Grade A, MAP = Grade A, SBP = Grade B<br>AAMI Standard met for: DBP, MAP<br>MAE: DBP = 3.449 mmHg, MAP = 2.310 mmHg, SBP = 5.727 mmHg |

Note: Shaded articles used deep learning and have cited our article in the archive. Athaya et al. [47] used the concept of our work but did not cite our work. Please refer to the arxiv version of this article [69].

## 5. Conclusion

In this work, we have tried to infer the complete waveform of blood pressure signals using PPG signals alone. There is indeed a plethora of works where only the various information of interest like DBP, SBP, and MAP are computed. However, these studies have been constrained as they could not manage to represent the overall picture of the blood pressure, i.e., the complete waveform. Moreover, often they required additional signals like ECG to aid in computation. Furthermore, several of the algorithms compute some handcrafted features from the signals, which necessitates the use of perfectly shaped signals, free from noises and artifacts. Thus, in addition to exhibiting better performance in DBP, SBP, and MAP prediction despite not being explicitly trained to do so, our work advances the state of the art from two different dimensions. Firstly, it overcomes the restriction of using ideal signals only, and secondly, it makes the involvement of additional signals redundant. Since deep learning models, upon analysis of the data, compute high-level abstract features adaptively from the data, this alleviates the need for computation of manual handcrafted features which may impose additional criteria. In addition, since the very beginning, we have been motivated to develop methods based on the PPG signal alone. These considerations paved the way for PPG2ABP, a method to infer the continuous arterial blood pressure waveform. The resulting waveforms generated from PPG2ABP closely resemble the actual waveform of the blood pressure signal, retaining the shape, amplitude, and even phase. However, the success of PPG2ABP extends beyond that. As has been mentioned earlier, there exists a plethora of methods to compute measures like SBP, DBP, and MAP. It turns out that, we can calculate these values using the pressure waveform estimated by PPG2ABP with outstanding accuracy. It is certainly remarkable that the accuracy of predicting these values, which was not our objective at all, still outperforms the existing methods, tailored toward predicting them in the first place. These methods are mostly supervised and trained explicitly to infer these values; on the other hand, our proposed PPG2ABP even without any explicit such training, predicts those better. The application of PPG2ABP can be manifold. In addition to predicting the typical systolic and diastolic blood pressure values, the complete profile of the blood pressure can be achieved. This will allow doctors to monitor the blood pressure of their patients continuously. Moreover, the trend and pattern of blood pressure can be mapped to user behaviour and activities, which may lead to insightful findings. Since nowadays all the smartwatches and fitness bands comprise a PPG sensor, applications based on PPG2ABP can be easily deployed to the mass market. This is due to the wonderful fact that PPG2ABP does not require any additional, expensive sensors like ECG. Therefore, to make the results of this research accessible to the general public, the codes have been made open source [63] and are released under the MIT License. Many future research directions may be explored. Firstly, we have used two networks, one to approximate and the other to refine the estimations. We are interested to design and develop a more optimized deep learning model that we will be able to replace the two with only one model instead. Furthermore, it will be interesting and beneficial to develop applications for wearables using PPG2ABP and conduct clinical studies thereby. Moreover, many of the existing methods have shown the effectiveness of utilizing a personalized calibration, which we too wish to explore in our future works.

**Acknowledgment**

This work was partially supported by NPRP12S-0227-190164 from the Qatar National Research Fund, a member of Qatar Foundation, a member of Qatar Foundation, Doha, Qatar. The statements made herein are solely the responsibility of the authors.

## Data and Code Availability

The data used in this experiment along with other relevant documents used to complete this work have been made available or updated in the following GitHub repository [63].